%% file: note_triangle_beta_s.tex
\documentclass[a4paper,11pt,numreference,epsfig,cite]{article}

\usepackage[latin1]{inputenc}
\usepackage{textcomp}
\usepackage{graphicx}
\usepackage{amsmath}

\newcommand{\stt}{\small\tt}

\newcommand{\Abstract}[1]{\begin{center}{\stt ABSTRACT}\end{center}{#1}
\vskip 0.4in}

\input{note_macro.tex}

\title{ \qquad CP violation and determination of the $bs$ \qquad  "flat'' unitarity triangle at FCCee\protect\\}

\author{R. Aleksan$^1$, L. Oliver$^2$ and E. Perez$^3$ \\
\footnotesize $^1$ IRFU, CEA, Universit\'e Paris-Saclay, 91191 Gif-sur-Yvette cedex, France \\
\footnotesize $^2$ IJCLab, P\^ole Th\'eorie, CNRS/IN2P3 et Universit\'e Paris-Saclay, b\^at. 210, 91405 Orsay, France \\
\footnotesize $^3$ CERN, EP Department, Geneva, Switzerland}

\begin{document}

\maketitle

\Abstract{\normalsize \baselineskip 22pt
We investigate the sensitivity with which two angles of the ``flat'' unitarity triangle, defined by $ V_{ub}^* V_{us}  + V_{cb}^* V_{cs}  + V_{tb}^* V_{ts} =0$, can possibly be measured directly at FCCee. We show that the measured errors on the angle $\alpha_s = \arg(- V_{ub}^* V_{us} /V_{tb}^* V_{ts})$ and $\beta_s =  \arg(-V_{tb}^* V_{ts} / V_{cb}^* V_{cs})$ should be better than $0.4^\circ$ and $0.035^\circ$, respectively. These measurements, combined with the measurement of the 3rd angle $\gamma_s = \arg(- V_{cb}^* V_{cs}/ V_{ub}^* V_{us})$, discussed in a different paper, will contribute to probe further the consistency of the CP sector of the Standard Model with unprecedented level of accuracy.}

 
\section{Introduction}
    
\noindent The purpose of this note is to study with what accuracy can the angles of a ``non usual''  Unitarity Triangle be determined directly at FCCee. Indeed, in most of the publications, the authors concentrate on the ``so called'' Unitarity Triangle, one of whose angles is obtained from the ``golden'' channel $B^0 (\overline{B^0})\to J/\psi K_s$. This triangle has its three sides of the same order of magnitude and is therefore the less difficult to measure. However, there are five other triangles, two of which are rather ``flat'', and the other two are ``very flat''.
In order to test more thoroughly the Standard Model (SM), it would be interesting to measure these other triangles directly and independently. By {\it directly} we mean the measurement of an angle without making use of its relations to other angles in the SM. In this note we investigate the sensitivity which one may expect at FCCee~\cite{fccee:1,fccee:2,fccee:3}.

 In the SM, one derives the unitarity relations from the CKM quark mixing matrix~\cite{CKM:1},
\be
V_{CKM} =
\begin{bmatrix}
V_{ud} & V_{us} & V_{ub} \\
V_{cd} & V_{cs} & V_{cb} \\
V_{td} & V_{ts} & V_{tb} \\
\end{bmatrix}
\label{eq:CKMmatrix}
\ee

Should there be only 3 families of quarks, the unitarity relations, which are derived from $V_{CKM}V_{CKM}^\dagger=1$, read as below:
\be
\begin{array}{cccccl} 
UT_{db} & \equiv & V_{ub}^* V_{ud}  + V_{cb}^* V_{cd}  + V_{tb}^* V_{td} & = & 0 \\
UT_{sb} & \equiv & V_{ub}^* V_{us}  + V_{cb}^* V_{cs}  + V_{tb}^* V_{ts} & = & 0 \\
UT_{ds} & \equiv & V_{us}^* V_{ud}  + V_{cs}^* V_{cd}  + V_{ts}^* V_{td} & = & 0 \\
\end{array}
\label{eq:UT}
\ee

\noindent plus three additional triangular relations. The three relations of Equation~(\ref{eq:UT}) are visualized in Figure 1: 
\begin{center}
\includegraphics[scale=0.6]{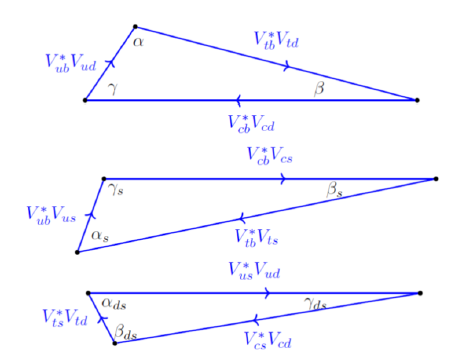} \qquad\\
\end{center}
Fig. 1: Unitarity Triangle UT$_{db}$ involving the $1^{st}$ and $3^{rd}$ columns (top),  Unitarity Triangle UT$_{sb}$ involving the $2^{nd}$ and $3^{rd}$ columns (center) and Unitarity Triangle UT$_{ds}$ involving the $1^{st}$ and $2^{nd}$ columns (bottom) of the CKM matrix. Note that these triangles are not to scale.

\vskip 15pt

In the SM, the CKM matrix has only 4 independent parameters. Therefore the angles of these triangles can be expressed in terms of four  angles~\cite{akl:1}. The first relation in Equation~(\ref{eq:UT}) is known as the Unitarity Triangle, with the three sides of the same order, and has been studied extensively. However the other ones would deserve to be studied in detail as well, in order to investigate further the consistency of the SM. 

We define the angles of these triangles as 

\be
\begin{array}{cccccl} 
\alpha =  \arg \left(-\frac{  V_{tb}^* V_{td} }{ V_{ub}^* V_{ud} }\right)    ,   
\beta =   \arg \left(-{ V_{cb}^* V_{cd} \over V_{tb}^* V_{td} }\right) , 
\gamma =  \arg \left(-{  V_{ub}^* V_{ud} \over V_{cb}^* V_{cd} }\right)  \\
 \end{array}
\label{eq:UT6-angles}
\ee
\be
\begin{array}{cccccl} 
\alpha_s =  \arg \left(-\frac{  V_{ub}^* V_{us} }{ V_{tb}^* V_{ts} }\right)    ,   
\beta_s =   \arg \left(-{ V_{tb}^* V_{ts} \over V_{cb}^* V_{cs} }\right) , 
\gamma_s =  \arg \left(-{  V_{cb}^* V_{cs} \over V_{ub}^* V_{us} }\right)  \\
\end{array}
\label{eq:UT5-angles}
\ee
\be
\begin{array}{cccccl} 
\alpha_{ds} =  \arg \left(-\frac{  V_{us}^* V_{ud} }{ V_{ts}^* V_{td} }\right)    ,   
\beta_{ds}  =  \arg \left(-{  V_{ts}^* V_{td} \over V_{cs}^* V_{cd} }\right)  , 
\gamma_{ds} =   \arg \left(-{ V_{cs}^* V_{cd} \over V_{us}^* V_{ud} }\right) \\
\end{array}
\label{eq:UT4-angles}
\ee

\vskip 7pt

For the angles $\alpha , \ \beta , \ \gamma$ of the triangle UT$_{db}$ we have adopted the usual convention with circular permutation of the quarks $t, \ c, \ u$. 
For the triangle UT$_{sb}$, we have adopted the generally accepted notation for $\beta_s$ ~\cite{beta_s}, and for the other angles the corresponding circular permutations of $t, \ c, \ u$. For the triangle UT$_{ds}$, we have used the notation of UT$_{sb}$ with the replacement $b \to s, \ s\to d$.

\vskip 10pt

To have a feeling on how these different triangles compare in the CKM scheme, let us give the angles of $UT_{sb}$ and $UT_{sd}$ using the PDG (2020) improved Wolfenstein parametrization with the central values for the relevant parameters $\lambda \simeq 0.226, A \simeq 0.790, \eta \simeq 0.357, \rho \simeq 0.141$ \cite{PDG 2020}, that fit the ordinary unitarity triangle $UT_{db}$, 

$$UT_{sb} : \ \qquad \qquad (\alpha_s, \beta_s, \gamma_s) \simeq (1.177, 0.018, 1.947)$$
\be
UT_{ds} : \ \ \  (\alpha_{ds}, \beta_{ds}, \gamma_{ds}) \simeq (0.402, 2.748, 5.8 \times 10^{-4})
\label{eq:Values of angles CKM}
\ee

\vskip 8pt

In this paper we concentrate on the Triangle  UT$_{sb}$, which is rather ``flat'' since two of its sides are of order $\lambda^2$, where $\lambda = \sin\theta_c\simeq 0.22$, while the 3$^{rd}$ side is of order $\lambda^4$. The three angles of this triangle can be measured directly at FCCee. The angle $\beta_s$ can be measured with precision through the well-known, and already widely used, final states $\overline{B}_s(B_s) \to J/\psi \phi$ or $J/\psi \eta$. Once this angle is known, one could measure the angle $\alpha_s$ by the final states $\overline{B}_s (B_s) \to D_s^+K^-$, $D^0\phi$ and their CP conjugated modes $B_s(\overline{B}_s) \to D_s^-K^+$, $\overline{D}^0\phi $, as will be discussed below.\par

Finally, the decays $B^\pm \to \overline{D}^0(D^0)K^\pm$ (and their CP conjugated) can determine the angle $\gamma_s$, as will be discussed in a separate paper \cite{Aleksan-Oliver}.

\section{Study of interference effect for $\overline{B_s}(B_s)$ induced through the interplay of mixing and decay}

\subsection{$B_s$ decays to non-CP eigenstates}

\noindent We first investigate the sensitivity for measuring CP violation at FCCee with the modes $\overline{B_s} (B_s) \to D_s^+K^-$, $D^0\phi$ and their CP conjugates (see Figs. 2 and 3), that measure the same CP violating phase~\cite{adk:1}.

\includegraphics[scale=0.8]{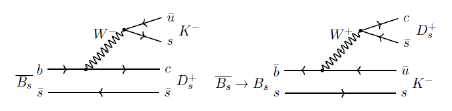} \qquad \\
Fig. 2: The leading Feynman diagrams for the $\overline{B_s}$ decay to the final state $D^+_sK^-$. There are also exchange diagrams, which are expected to be small and involve the same CKM elements.

\vskip 10pt

\includegraphics[scale=0.8]{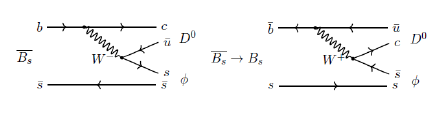} \qquad\\
Fig. 3: The leading Feynman diagrams for the $\overline{B_s}$ decay for the final state $D^0\phi$. There are also exchange diagrams, which are expected to be small and involve the same CKM elements. 

\vskip 10pt

 The eigenstates of the $2 \times 2$ $B_s$ mass matrix $M - {i \over 2} \Gamma$ are written in terms of the $B_s$ and $\overline{B_s}$ states as follows :
\be
\begin{array}{ccccl} 
|B_{L(H)}>& = & p|B_s> + (-) q|\overline{B_s}>  \\
\end{array}
\label{eq:Bs_LH}
\ee

\noindent In the Standard Model the box diagrams in Fig. 4, which are responsible for $B_s - \overline{B_s} $ mixing, dominated by $t$-quark exchange. 

\begin{center}

\includegraphics[scale=0.8]{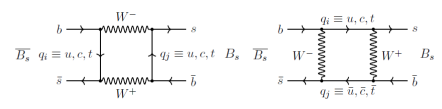} \qquad\\

\end{center} \par

Fig 4: The box Feyman diagrams for the $\overline{B_s}-B_s$ mixing, dominated by $t-$quark exchange in the SM.

\vskip 10pt

\noindent Thus one can safely use the approximation $\left| q / p \right| \simeq 1$, where $q/p$ is given by a ratio of CKM matrix elements $V_{tb}^* V_{ts}$. This approximation is good at the sub per mille level,

\be
\begin{array}{ccccl} 
\left({q \over p} \right)_{B_s} & = &- \sqrt{ {M^*_{12} \over M_{12} } }& \simeq & - {V_{tb}^* V_{ts} \over V_{tb} V_{ts}^*} \\
\end{array}
\label{eq:Bs_mixing_2}
\ee

\noindent Similarly the phase of $q/p$ is given by $\arg(- V_{tb}^* V_{ts}/V_{tb} V_{ts}^*)$ to a sub per mille level as well. Note that the quantity above is not invariant under phase convention and thus is not an observable. Let us now write also 
 \be
\begin{array}{ccccccc} 
\lambda (f) &=& {q \over p} { < f |\overline{B_s} > \over < f |B_s > } & , & \overline{\lambda} (\overline{f}) &=& {p \over q} { <\overline{f} |{B_s} > \over < \overline{f} |\overline{B_s} > } \\
\lambda (\overline{f}) &=& {q \over p} { < \overline{f} |\overline{B_s} > \over < \overline{f}|B_s > } & , & \overline{\lambda} (f) &=& {p \over q} { < f |{B_s} > \over < f |\overline{B_s} > } \\
\end{array}
\label{eq:Bs_lambda_1}
\ee
and with the notation $\rho =\left|\lambda(f) \right|$, one has 
\be
\begin{array}{ccccccc} 
\lambda (f) &=& \rho e^{i(\phi_{CKM} +\delta_s)} & , &  \overline{\lambda} (\overline{f}) &=& \rho e^{i(-\phi_{CKM} +\delta_s)}\\
\lambda (\overline{f}) &=& {1 \over \rho} e^{i(\phi_{CKM} -\delta_s)} & , &  \overline{\lambda} (f) &=& {1 \over \rho} e^{-i(\phi_{CKM} +\delta_s)}\\
\end{array}
\label{eq:Bs_lambda_1}
\ee

\noindent where $\phi_{CKM}$ and $\delta_s$ are the CKM phase difference and strong phase difference of e.g. the two diagrams in Fig. 2, respectively.\par 
Some words of caution. Note that equations~(\ref{eq:Bs_lambda_1}) are relevant only for $\rho \neq 0$. The modes $\overline{B_s}(B_s) \to D_s^+K^-$ and their CP conjugated are in this case, as can be seen from the two diagrams of Fig. 2. However, for the pionic mode $\overline{B_s} \to D_s^+\pi^-$ and its CP conjugate, which are useful to measure the mixing parameter and wrong tagging fraction, as
will be discussed below, it is not possible to get the same final state through $\overline{B_s} - B_s$ mixing, and only the decays $< \overline{f} \mid \overline{B_s} >$ and $< f \mid B_s >$ are possible.

\vskip 10pt

\noindent The complete time-dependent decay widths are 
\be
\begin{array}{ccl} 
\Gamma (B_s(t) \to f ) &  = & \mid < f \mid B_s > \mid^2\  e^{-\Gamma t}  \{ {1+\rho^2\over 2}\cosh {\Delta \Gamma t\over 2} + {1-\rho^2\over 2}\cos\Delta m t   \\
& & +\ \rho \cos \phi_{CP}^+\sinh {\Delta \Gamma t\over 2} -\  \rho \sin \phi_{CP}^+ \sin \Delta m t \} \\
\Gamma (\overline{B_s}(t) \to f ) &  = & \mid < f \mid B_s > \mid^2\ e^{-\Gamma t}  \{ {1+\rho^2\over 2}\cosh {\Delta \Gamma t\over 2} - {1-\rho^2\over 2}\cos\Delta m t   \\
& & +\ \rho \cos \phi_{CP}^+\sinh {\Delta \Gamma t\over 2} +\  \rho \sin \phi_{CP}^+ \sin \Delta m t \} \\
\Gamma (B_s(t) \to \overline{f} ) &  = & \mid < f \mid B_s > \mid^2\  e^{-\Gamma t}  \{ {1+\rho^2\over 2}\cosh {\Delta \Gamma t\over 2} - {1-\rho^2\over 2}\cos\Delta m t   \\
& & +\  \rho \cos \phi_{CP}^-\sinh {\Delta \Gamma t\over 2} -\  \rho \sin \phi_{CP}^- \sin \Delta m t \} \\
\Gamma (\overline{B_s}(t) \to \overline{f} ) &  = & \mid < f \mid B_s > \mid^2\  e^{-\Gamma t}  \{ {1+\rho^2\over 2}\cosh {\Delta \Gamma t\over 2} + {1-\rho^2\over 2}\cos\Delta m t   \\
& & +\ \rho \cos \phi_{CP}^-\sinh {\Delta \Gamma t\over 2} +\  \rho \sin \phi_{CP}^- \sin \Delta m t \} \\
\end{array}
\label{eq:Bs_decay_full}
\ee
\noindent In eqn. (\ref{eq:Bs_decay_full}), terms of order $a = {\rm Im} {\Gamma_{12} \over m_{12}}$ have been neglected ($\mid q/p \mid^2 = 1 - a$), $\rho =\left|\lambda(f) \right|$ is the modulus of the ratio of the amplitudes (\ref{eq:Bs_lambda_1}), $\Delta m = m_H - m_L \simeq \ 17.757\ {\rm ps}^{-1}$ is the mass difference between the $B_s$ eigenstates ($H, L$ stand for heavy and light states), $\Gamma = {\Gamma_H + \Gamma_L \over 2}$, $\Delta \Gamma = \Gamma_L - \Gamma_H$ (so that $\Delta \Gamma > 0$ for $K$ and $B_s$ mesons), and $\phi_{CP}^\pm = \phi_{CKM} \pm \delta_s$, where $\phi_{CKM}$ is the CP-violating weak phase and $\delta_s$ is the difference between the strong phases of the interfering diagrams (see for example Fig. 2). Moreover, if the time dependence is normalized to an integrated finite interval in $t$ that is not $[ 0,\infty]$, one should introduce a normalization factor $N_f$ in (\ref{eq:Bs_decay_full}).\par
In the following, for the sake of simplicity, we neglect the width difference $\Delta \Gamma_s = (0.090 \pm 0.005)\times 10^{12}\ s^{-1}$ \cite{PDG 2020}, the effect of which is negligible in this study. Note however that these terms with  $\Delta \Gamma_s$ help to remove ambiguities for the extraction of $\phi_{CKM}$.\par

\vskip 3truemm
In order to observe these distributions experimentally, one needs to tag the nature ($B_s$ or $\overline{B}_s$) of the initial $B$-meson. Unfortunately, this tagging is imperfect, and one needs to introduce the wrong tagging fraction $\omega$ in the equation above. We define thus the experimental distributions :
$$\Gamma(B_s \to f)_{exp} = (1-\omega)\Gamma(B_s \to f) + \omega\Gamma(\overline{B}_s \to f)$$
$$\Gamma(\overline{B}_s \to f)_{exp} = (1-\omega)\Gamma(\overline{B}_s \to f) + \omega\Gamma(B_s \to f)$$
\be
\Gamma(B_s \to \overline{f})_{exp} = (1-\omega)\Gamma(B_s \to \overline{f}) + \omega\Gamma(\overline{B}_s \to \overline{f})
\label{eq:wrong_tagging}
\ee
$$\Gamma(\overline{B}_s \to \overline{f})_{exp} = (1-\omega)\Gamma(\overline{B}_s \to \overline{f}) + \omega\Gamma(B_s \to \overline{f})$$

\vskip 3truemm
\noindent Including the wrong tagging fraction $\omega$, equations~(\ref{eq:Bs_decay_full}) can thus be approximated as 
\be
\begin{array}{ccl} 
\Gamma (B_s(t) \to f ) &  = & \mid < f \mid B_s > \mid^2 \ e^{-\Gamma t}  \{ [1-\omega (1-\rho^2)] \cos^2 {\Delta m t \over 2}  \\
& & +\ [\rho^2 +\omega (1-\rho^2)] \sin^2 {\Delta m t \over 2} \\
& & -\ (1-2\omega ) \rho \sin \phi_{CP}^+ \sin \Delta m t \} \\
\Gamma (\overline{B_s}(t) \to f ) &  = & \mid < f \mid B_s > \mid^2\ e^{-\Gamma t}  \{ [\rho^2 +\omega (1-\rho^2)] \cos^2 {\Delta m t \over 2}  \\
& & +\ [1-\omega (1-\rho^2)] \sin^2 {\Delta m t \over 2} \\
& & +\ (1-2\omega ) \rho \sin \phi_{CP}^+ \sin \Delta m t \} \\
\Gamma (B_s(t) \to \overline{f}) &  = & \mid < f \mid B_s > \mid^2\ e^{-\Gamma t}  \{ [\rho^2 +\omega (1-\rho^2)] \cos^2 {\Delta m t \over 2}  \\
& & +\ [1-\omega (1-\rho^2)] \sin^2 {\Delta m t \over 2} \\
& & -\ (1-2\omega ) \rho \sin \phi_{CP}^- \sin \Delta m t \} \\
\Gamma (\overline{B_s}(t) \to \overline{f} ) &  = &\mid < f \mid B_s > \mid^2\ e^{-\Gamma t}  \{ [1-\omega (1-\rho^2)] \cos^2 {\Delta m t \over 2}  \\
& & +\ [\rho^2 +\omega (1-\rho^2)] \sin^2 {\Delta m t \over 2} \\
& & +\ (1-2\omega ) \rho \sin \phi_{CP}^- \sin \Delta m t \} \\

\end{array}
\label{eq:Bs_decay}
\ee

One notes that the term including the CP violating phase $\phi^\pm_{CP}$ vanishes when $\omega = 0.5$. It is thus important to develop a tagging algorithm reducing $\omega$ as much as possible. As will be discussed later, for CP eigenstates $\rho = 1$ and $\delta_s = 0$. If $\rho = 0$ as for $\overline{B}_s \to D_s^+ \pi^-$, one finds the equations of the sole mixing.

\vskip 10pt
Using all four decays in equation~(\ref{eq:Bs_decay}), one can extract $\sin^2\phi_{CKM}$ with a 2-fold ambiguity:
\be
\sin^2\phi_{CKM} = {1+\sin\phi^+_{CP} \sin\phi^-_{CP}   \pm \sqrt{  (1-\sin\phi^{+\ \ 2}_{CP}) (1-\sin\phi^{-\ \ 2}_{CP}) } \over 2} 
\label{eq:Bs_sinphi BIS}
\ee

\noindent where $\phi_{CKM} = \pi - \alpha_s + \beta_s = \pi -\gamma + 2 \beta_s - \gamma_{ds}$.

\subsection{$B_s$ decays to CP eigenstates}

For CP eigenstates one has $\rho = 1$ and $\delta_s =0$ and one has therefore,
\be
\begin{array}{ccl} 
\Gamma (B_s(t) \to f_{CP} ) &  = & | < f_{CP} |B_s >|^2 \ e^{-\Gamma t}  \{ 1
 -\ (1-2\omega ) \eta_f \sin \phi_{CP} \sin \Delta m t \} \\
\Gamma (\overline{B_s}(t) \to f_{CP} ) &  = & | < f_{CP} |B_s >|^2 \ e^{-\Gamma t}  \{ 1+\ (1-2\omega ) \eta_f\sin \phi_{CP} \sin \Delta m t \} \\

\end{array}
\label{eq:Bs_CPdecay}
\ee
\noindent where $\eta_f$ is the CP eigenvalue of the final state, e.g. $\eta_f = +1$ for $J/\psi \eta$. 

\subsection{Strategy to determine the angles $\beta_s$ and $\alpha_s$}

\subsubsection{CP-eigenstates}

\noindent Let us now investigate the CP eigenstate final state modes such as $\overline{B_s}(B_s) \to J/\psi \phi$ (for definite longitudinal, parallel or perpendicular transversity polarizations) or $J/\psi \eta$ (see Fig. 5). 

\vskip 10pt

\includegraphics[scale=0.8]{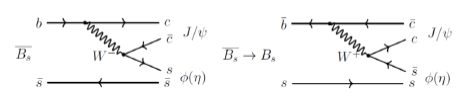} \qquad \qquad \\
Fig. 5: The leading Feynman diagrams for the $\overline{B_s}$ decay for the final state $J/\psi \phi$ or $J/\psi \eta$. There are also exchange diagrams, which are expected to be small and involve the same CKM elements. Penguin diagrams also exist and include other CKM elements, but they are expected to be negligible.

\vskip 10pt

The angle $\beta_s$ can be determined by the angular analysis of $\overline{B}_s (B_s) \to J/\psi \phi \to \mu^+\mu^- K^+ K^-$ \cite{Dighe-Dunietz-Fleischer}. It can also be measured by other decay modes where no angular analysis is necessary, like $\overline{B}_s (B_s) \to J/\psi \eta$. 

For a V$_1$-V$_2$ final state such as $J/\psi \phi$, $\eta_f$ depends of the polarization state. For $J/\psi \phi$, it is $\eta_f=+1$ for the longitudinal and parallel polarizations while it is $\eta_f=-1$ for the perpendicular polarization. In case one does not disentangle the polarization by doing a full angular analysis, one would have an effective $\eta_f=1-2f_\perp = 0.510\pm0.008$ as obtained from the experimentally measured polarizations~\cite{PDG 2020} summarized in Table 1.

\vskip 10pt

$$ \begin{tabular}{cccc}
\hline
& $f_L = \Gamma_L/\Gamma$ & $f_\parallel= \Gamma_\parallel/\Gamma $ & $f_\perp= \Gamma_\perp/\Gamma$  \\
\hline\hline
$CP$ & $+1$ & $+1$ & $-1$\\
$J/\psi \phi$ & $0.527 \pm 0.008$ & $0.228 \pm 0.007$ & $0.245\pm 0.004$\\
\hline
\end{tabular}$$

\vskip 5pt

Table 1: Polarization fractions of the final state $J/\psi \phi$. It is interesting to note that due to the $V-A$ structure of the SM model, one expects the helicity amplitude $A_+ \simeq 0$, and therefore the relation between transversity fractions $f_\parallel \simeq f_\perp$, in agreement with the measured values within 2$\sigma$.

\vskip 10pt

For this mode one gets the product of the CKM elements,
\be
- {V_{tb}^* V_{ts} \over V_{tb} V_{ts}^*} {V_{cb} V_{cs}^* \over V_{cb}^* V_{cs}} = - {V_{tb}^* V_{ts} \over V_{cb}^* V_{cs}} {V_{cb} V_{cs}^* \over V_{tb} V_{ts}^*} = {\mid V_{cb} V_{cs}^*\mid^2 \over \mid V_{tb} V_{ts}^*\mid^2}\ e^{i(\pi + 2 \beta_s)}
\label{eq:CKM_Bs_CP}
\ee

\noindent and therefore one obtains
\be
\phi_{CP}(J/\psi \phi) = \pi + 2 \beta_s
\label{eq:phi_1}
\ee
From $\overline{B}_s (B_s) \to J/\psi \phi$, LHCb gives the present precision on this angle by the direct measure $\phi_s^{c\overline{c}s} = -2 \beta_s = -0.083 \pm 0.041$ \cite{LHCb J/Psi phi}, compared to the SM model prediction $\beta_s = 0.0185 \pm 0.0003$ \cite{CKMfitter Descotes}. It is worth to emphasize that at present $\beta_s$ is poorly measured, and that the future experiment FCCee would give a much more precise determination.

\subsubsection{Non-CP eigenstates}

For $\overline{B_s} \to D_s^+K^-, D^0\phi$, from (\ref{eq:Bs_mixing_2}), one gets the invariant product of CKM elements,
\be
\begin{array}{ccccl} 
\lambda( D_s^+K^-) = -  {V_{tb}^* V_{ts} \over V_{tb} V_{ts}^*} \times {V_{cb} V_{us}^* \over V_{ub}^* V_{cs}} & =
\left|{V_{cb} V_{us}^* \over V_{tb} V_{ts}^*}\right| \times \left|{V_{tb}^* V_{ts} \over V_{ub}^* V_{cs}}\right| e^{i\phi_{CKM}}\\
\end{array}
\label{eq:CKM_Bs_1}
\ee

\vskip 10pt

\noindent where $\phi_{CKM}$ is the weak phase violating CP. One can rewrite the expression above as :
\be
\lambda( D_s^+K^-) = - {V_{tb}^* V_{ts} \over V_{ub}^* V_{cs}} \times {V_{cb} V_{us}^* \over V_{tb} V_{ts}^*} = \left| {V_{cb}  \over V_{ub}} \right|^2 \left|{V_{ub} V_{us}^* \over V_{cb}^* V_{cs}} \right| e^{i(\pi -\alpha_s +\beta_s)}
\label{eq:CKM_Bs_2}
\ee

\noindent and therefore one gets 
\be
\phi_{CKM}(D_sK) = \pi -(\alpha_s-\beta_s)
\label{eq:CKM_Bs_2bis}
\ee

\noindent where $\alpha_s$ and $\beta_s$ are angles of the 2$^{nd}$ Unitarity Triangle in Fig. 1. If $\beta_s$ is known from the angular analysis of $\overline{B}_s \to J/\psi \phi \to \mu^+ \mu^- K^+ K^-$, eqn. (\ref{eq:CKM_Bs_2bis}) shows that the modes $D_s K$ would allow a {\it direct determination} of $\alpha_s$.

\vskip 10pt

Since in the SM the sum of the three angles is $\pi$, one has $\phi_{CKM} = \pi -(\alpha_s-\beta_s) = 2 \beta_s + \gamma_s$, and one could have an estimation of $\gamma_s$ although this will not be a direct determination since the SM has been assumed. This is a similar situation to the angle $\gamma$ of the ordinary unitarity triangle $UT_{db}$. It was proposed by I. Dunietz to measure the combination $-(2\beta + \gamma)$ by the decay modes $\overline{B}_d(B_d) \to D^+\pi^-$ and their CP conjugated ~\cite{Dunietz}. The angle $\beta$ was well-known at the time, and this combination could give information on the still poorly determinated angle $\gamma$.

\vskip 10pt

Therefore, once one obtains a measurement of the angle $\beta_s$ from CP-eigenstates, by using relation (\ref{eq:CKM_Bs_2bis}) one could get information on the angle $\alpha_s$ from the time-dependent CP asymmetries for $\overline{B_s} \to D_s^+K^-, D^0\phi$ or $\overline{B_s} \to D_s^-K^+, \overline{D}^0\phi$ and their CP-conjugated modes.
On the other hand, since the angle $\beta_s$ is small, a measurement of $\phi_{CKM}$ (\ref{eq:CKM_Bs_2bis}) would give a rather good preliminary determination of $\alpha_s$.  

\vskip 10pt

Let us now look at relations of $\phi_{CKM}$ to other angles, in particular those of the standard $T_{db}$ triangle. This is possible because the number of independent parameters of the CKM matrix is four, as exposed in ref. \cite{akl:1}.\par

Indeed, one can write  $\alpha_s$ as

\be
\begin{array}{llllll} 
\alpha_s & = & \arg\left(-{V_{ub}^* V_{us} \over V_{tb}^* V_{ts}}\right) = \arg \left( - {V_{ub}^* V_{us} \over V_{tb}^* V_{ts}} \times { V_{ud}V_{cb}^*V_{cd} V_{cs}\over V_{ud}V_{cb}^*V_{cd}V_{cs}  } \right)\\ 
& = & \arg\left( - {V_{ub}^* V_{ud} \over V_{cb}^* V_{cd}}\right) -
\arg \left( - {V_{tb}^* V_{ts} \over V_{cb}^* V_{cs}} \right) +
\arg \left( - { V_{cs}^*V_{cd} \over  V_{us}^*V_{ud}} \right) \\
& = & \gamma - \beta_s + \gamma_{ds}
\end{array}
\label{eq:CKM_Bs_3}
\ee

\noindent where $\gamma$ is one of the angles of the usual Unitarity Triangle involving  the $1^{st}$ and  $3^{rd}$ columns of the CKM matrix and $\gamma_{ds}$ corresponds to one of the angles of the 3rd triangle in equation~(\ref{eq:UT}) (see Fig. 1), that is very small. Hence one could approximate $\alpha_s$ by
\be
\begin{array}{cccccc} 
\alpha_s \simeq \gamma -\beta_s\\
\end{array}
\label{eq:alpha_s}
\ee

\noindent and from relations (\ref{eq:CKM_Bs_2bis}-\ref{eq:alpha_s}) one gets, neglecting $\gamma_{ds}$,
\be
\begin{array}{cccccc} 
\phi_{CKM} \simeq \pi - \gamma  +2\beta_s\\
\end{array}
\label{eq:alpha_sbis}
\ee

\noindent a relation that would be an interesting test of the CKM scheme, since $\beta_s$ appears as a correction to an angle of the ordinary unitarity triangle $T_{db}$.\par
The determination of $\gamma$ using $B_s \to D_s K$ was first proposed by R. Aleksan et al. \cite{adk:1}. Relation (\ref{eq:alpha_sbis}) is known \cite{Nardulli}, and has been proposed at LHCb to measure the angle $\gamma$ once $\beta_s$ would be measured from the angular analysis of time-dependent CP violation 
in $\overline{B}_s(B_s) \to J/\psi \phi \to \mu^+ \mu^- K^+ K^-$.
\vskip 10pt

\section{Experimental expectations at FCCee}

\noindent As one can see in equations~(\ref{eq:Bs_decay}) and~(\ref{eq:Bs_CPdecay}), CP violating effects are damped by the fraction of wrong tagging $\omega$. Table 2 shows typical tagging performances of some experiments. It is thus essential to measure this factor precisely. Fortunately, the decay $\overline{B_s}\to D^+_s \pi^-$ allows one to measure the value of $\omega$ very precisely. 

$$ \begin{tabular}{cccc}
\hline
Tagging Merit & LEP & BaBar & LHCb  \\
\hline\hline
$\epsilon(1-2\omega)^2$ & 25-30\% & ~30\% & ~6\%\\
\hline
\end{tabular} $$
Table 2: Typical tagging Figure of Merit for some experiments, where $\epsilon$ is the tagging efficiency and $\omega$ the wrong tagging fraction, in the range $0-0.5$.

\subsection{Generic detector resolutions}

\noindent In order to carry out experimental studies, we define a generic detector, the resolution of which is  parametrized as follows :
\be
\begin{array}{lccl} 
\mathrm{Acceptance :}& |\cos \theta|&<&0.95\\
\hline
\mathrm{Charged \ particles :} & \\
\mathrm{ p_T\ resolution :}& {\sigma (p_T) \over p_T^2}  & = & 2. \times 10^{-5} \ \oplus \ {1.2 \times 10^{-3}\over p_T \sin \theta}\\
\mathrm{ \phi , \theta \ resolution :}& \mathrm{\sigma (\phi , \theta) \ \mu rad } & = &  18 \ \oplus \ {1.5 \times 10^{3} \over p_T\sqrt[3]{\sin \theta} }\\
\mathrm{Vertex \ resolution :}& \mathrm{\sigma (d_{Im}) \ \mu m} & = &  1.8 \ \oplus \ {5.4 \times 10^{1} \over p_T\sqrt{\sin \theta} }\\
\hline
\mathrm{e,\gamma \ particles :} & \\
\mathrm{Energy\ resolution:}&{\sigma (E) \over E}  & = & {5 \times 10^{-2} \over \sqrt{E}} \ \oplus \ 5 \times 10^{-3}\\
\mathrm{EM \ \phi , \theta \ resolution :}& \mathrm{\sigma (\phi , \theta) \ m rad } & = &  {7 \over \sqrt{E} }\\
\hline \hline
\end{array}
\label{eq:track_resolution}
\ee

\vskip 10pt

\noindent where $\theta, \phi $ are the particles' polar and azymutal angles respectively, $p_T$ (in GeV) the track transverse momentum, $E$ the $e^\pm,\gamma$ energy and $\mathrm{d_{Im}}$ the tracks' impact parameter.

\subsection{Measuring the wrong tagging fraction}

In order to extract experimentally the CKM phases using the time-dependent distributions in equations~(\ref{eq:Bs_decay}) and~(\ref{eq:Bs_CPdecay}), one needs to measure precisely the wrong tagging fraction. This can be done done by using the decay $\overline{B_s} \to D_s^+\pi^- \to (\phi \pi^+)_{D_s} \pi^- \to K^+K^-\pi^+\pi^-$. Indeed, only one diagram is involved in this decay (see Fig. 6) and thus no CP violation is expected. Furthermore its branching fraction is relatively large ($\sim3\times 10^{-3}$). Some $13.8\times 10^6$ $B_s + \overline{B_s}$ such decays are expected to be produced at FCCee when accumulating 150\ $ab^{-1}$ at the $Z$-pole (see Table 3 below). 

\begin{center}

\includegraphics[scale=0.8]{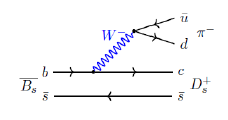} \qquad\\

\end{center}

Fig. 6: The Feynman diagrams for the $\overline{B_s}$ decay for the final state $D^+_s\pi^-$. This decay involves only one diagram, hence is very convenient for probing the $B_s$ tagging.

\vskip3truemm

\noindent One gets the time-dependent distributions from equation~(\ref{eq:Bs_decay}) with $\rho = 0$.
$$\Gamma (B_s(t) \to D_s^-\pi^+ ) =\ \mid <D_s^- \pi^+\mid B_s >\mid^2\  e^{-\Gamma t}  \{ (1-\omega) \cos^2 {\Delta m t \over 2}  +\ \omega \sin^2 {\Delta m t \over 2} \}$$
$$\Gamma (\overline{B_s}(t) \to D_s^-\pi^+ ) = \ \mid <D_s^- \pi^+\mid B_s >\mid^2\ e^{-\Gamma t}  \{ \omega \cos^2 {\Delta m t \over 2}  +\ (1-\omega) \sin^2 {\Delta m t \over 2} \}$$
$$\Gamma (B_s(t) \to D_s^+\pi^- ) = \ \mid <D_s^- \pi^+\mid B_s >\mid^2\ e^{-\Gamma t}  \{ \omega \cos^2 {\Delta m t \over 2}  +\ (1-\omega) \sin^2 {\Delta m t \over 2} \}$$
\be
\Gamma (\overline{B_s}(t) \to D_s^+\pi^- ) = \ \mid <D_s^- \pi^+\mid B_s >\mid^2\ e^{-\Gamma t}  \{ (1-\omega) \cos^2 {\Delta m t \over 2}  +\ \omega \sin^2 {\Delta m t \over 2} \}
\label{eq:Bs_time_Dspi}
\ee

The main source of background is the combinatorial one, which is however expected to be small, thanks to the excellent mass resolution on $D_s$ and $B_s$ as shown in Figure 7.

\includegraphics[scale=0.5]{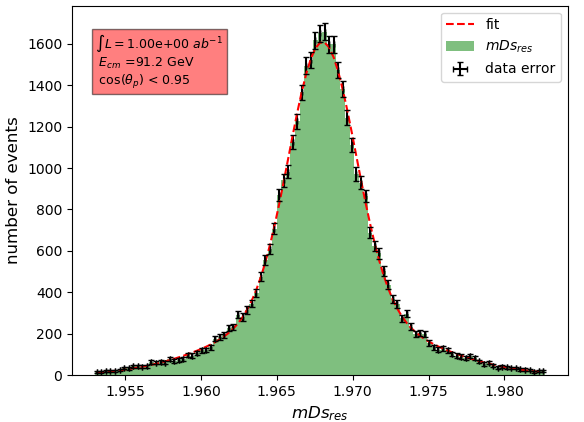} \\ 
\includegraphics[scale=0.5]{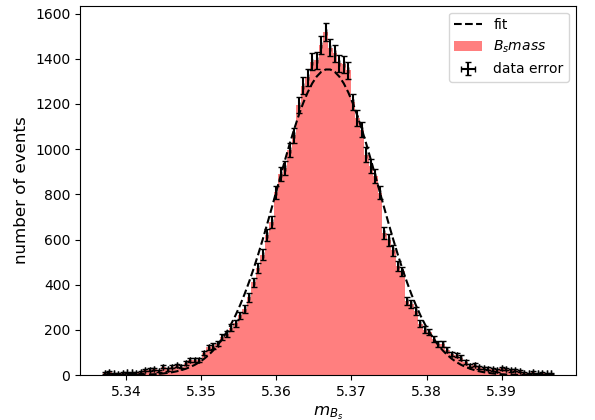} 

Fig. 7: $D_s$ (upper plot) and $B_s$ (lower plot) mass resolutions in $e^+ e^- \to Z \to \overline{B_s} (B_s) \to D_s^\pm \pi^\mp \to K^+K^- \pi^+ \pi^-$. One obtains $\sigma(m_{D_s}) \simeq 3.2$ MeV and  $\sigma(m_{B_s}) = 6.7$ MeV. The geometric acceptance of the detector ($|\cos \theta |< .95$) leads to an efficiency of 86$\%$ for this 4-body final state.

\includegraphics[scale=0.5]{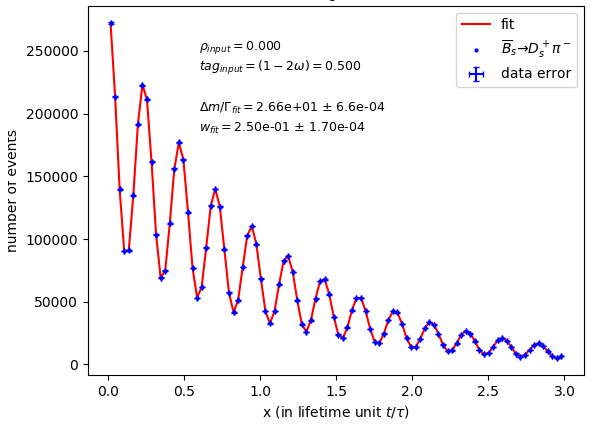} \qquad \qquad \qquad \qquad\\
\includegraphics[scale=0.5]{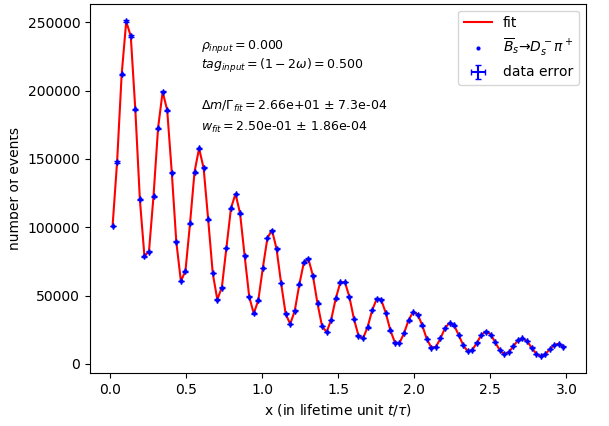}\par
Fig. 8: The time dependent distribution for $B_s \to D_s^-\pi^+$ + CP conjugate (upper plot) and $B_s \to D_s^+\pi^-$ + CP conjugate (lower plot). The overall statistical resolution on $\omega$ is $1.4\times 10^{-4}$.

\vskip5truemm

The expected time-dependent distributions corresponding to the equations~(\ref{eq:Bs_time_Dspi}) are shown in Fig. 8. One extracts the resolution of the wrong tagging parameter $\omega$ by fitting these time-dependent distributions. Thanks to the large statistics available, one finds $\sigma(\omega) \simeq 1.4\cdot 10^{-4}$. The oscillation frequency $\Delta m/\Gamma$ is also obtained with a resolution of about $5\cdot 10^{-4}$. In these figures, no resolution on the flight distance has been included. 
\begin{center}

\includegraphics[scale=0.5]{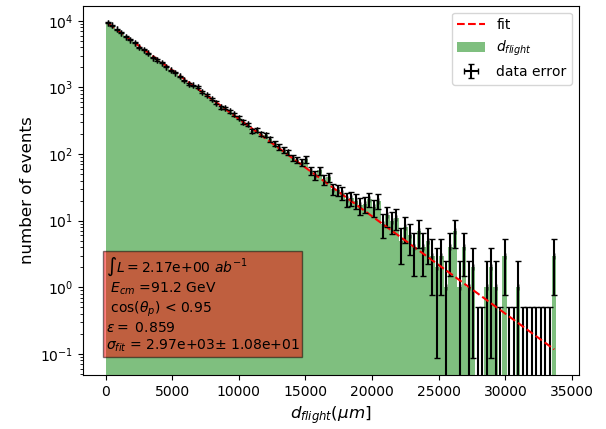}\\
\includegraphics[scale=0.5]{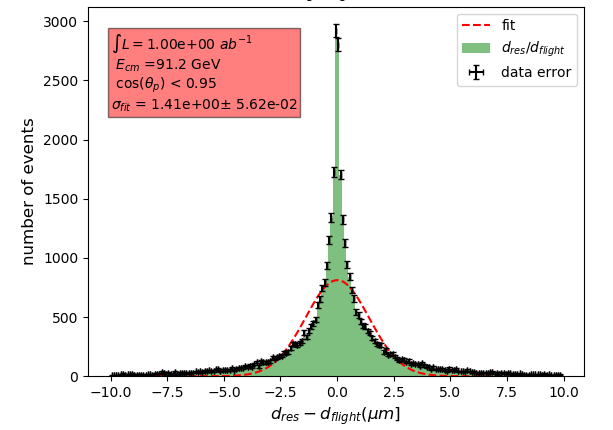}\\

\end{center}

Fig. 9: $B_s$ decay distance (upper plot) and resolution on decay distance due to momentum resolution (lower plot) in $e^+ e^- \to Z \to \overline{B_s} (B_s) \to D_s^\pm \pi^\mp \to K^+K^- \pi^+ \pi^-$. The uncertainty is  $~1.4\ \mu$m, and is negligible compared to the average flight distance of ~3\ mm.

\vskip5truemm

There are three sources of smearing of the $B_s$ oscillation frequency. 
\begin{enumerate}
\item The uncertainty on the $B_s$ reconstructed momentum: In Fig. 9 (upper plot) we show the $B_s$ flight distance. The average distance is $\sim 2.8$ mm. In Fig. 9 (lower plot) we show the resolution on flight distance due to the error on $B_s$ reconstructed momentum, which is negligible.
\item The uncertainty on primary vertex position: This resolution is obtained by a combination of the beam Interaction Point (IP) and reconstructed primary vertex. The error on the IP position at FCCee is obtained from the beam size: $\sigma (\mathrm{IP}_x) \simeq 4.5\ \mu$m, $\sigma (\mathrm{IP}_y) \simeq 0.02\ \mu$m and $\sigma (\mathrm{IP}_z)\simeq 0.30$\ mm. Combining these constraints with the measured tracks originating from the primary vertex results in an overall negligible primary vertex resolution.
\item The uncertainty on $B_s$ decay vertex position: This has been evaluated to be the dominant contribution. However, it is estimated that its impact on the $B_s$ flight distance is of the order of $\sim 18\ \mu$m (see Appendix). Overall, the resolution on the $B_s$ flight distance remains below $20\ \mu$m, and thus leads to a minor damping of the time dependent curves, which does not affect significantly the measurement of the CP violating angles.
\end{enumerate}

\subsection{Measurement of the CKM angles $\alpha_s$ and $\beta_s$}

We have now all the ingredients to evaluate the sensitivity on the CKM angles $\alpha_s$ and $\beta_s$ with the decays $\overline{B_s}(B_s )\to D_s^\pm K^\mp$ and $\overline{B_s}(B_s )\to J/\psi \phi$, respectively, which one expects at FCCee.

\vskip 10pt
Let us first concentrate on $\overline{B_s}(B_s )\to D_s^\pm K^\mp$. We simulate these reactions with the parametrized detector using some specific values for the strong phase difference, $\delta_s =40^\circ$, and the CKM phase, $\phi = 70^\circ$, which is very close to the expected value from the present indirect constraints on the angles of the corresponding unitarity triangle. 
We show in Fig. 10 the signal together with the main sources of background, with the expectation of the combinatorial one, which is however expected to be small, thanks to the excellent mass resolution.

As can be seen in Fig. 10, the signal is already rather clean without PID but becomes very clean once a simple PID is turned on. Using the signal, the time-dependent distributions corresponding to the equations~(\ref{eq:Bs_decay}) are shown in Fig. 11. From a global fit of these four time-dependent distributions, one extracts the resolution on the CKM phase $\Phi$ and the strong phase difference $\delta_s$. Thanks to the large statistics available, one finds $\sigma(\Phi) \simeq 0.4^\circ$.

\begin{center}

\includegraphics[scale=0.55]{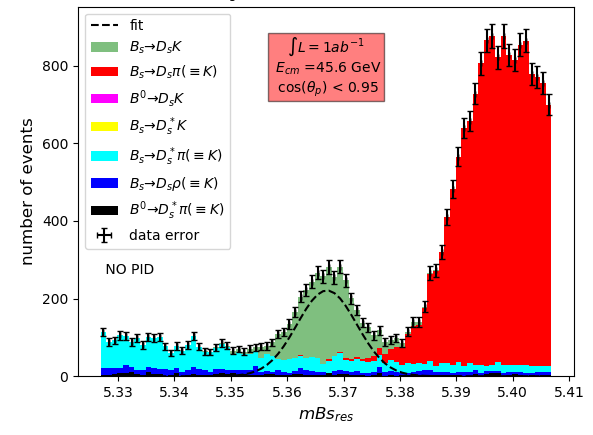} \qquad\\
\includegraphics[scale=0.55]{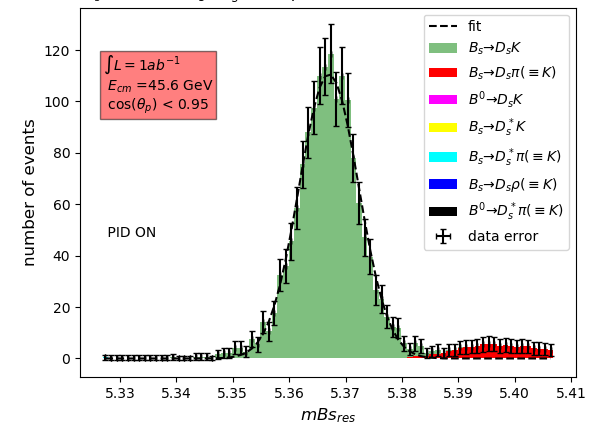} \qquad\\

\end{center} 

Fig. 10: $B_s$ mass resolutions in $\overline{B_s} (B_s) \to D_s^\pm K^\mp $ decays in $e^+ e^- \to Z \to \overline{B_s} (B_s) \to D_s^\pm K^\mp \to \phi \pi^\pm K^\mp \to K^+K^- \pi^\pm K^\mp$. The upper histogram shows the signal and backgrounds without particle identification. A simple particle identification (PID) based on Time of Flight (ToF) and $dE/dx$ is turned on for the lower histogram.

\begin{center}

\includegraphics[scale=0.55]{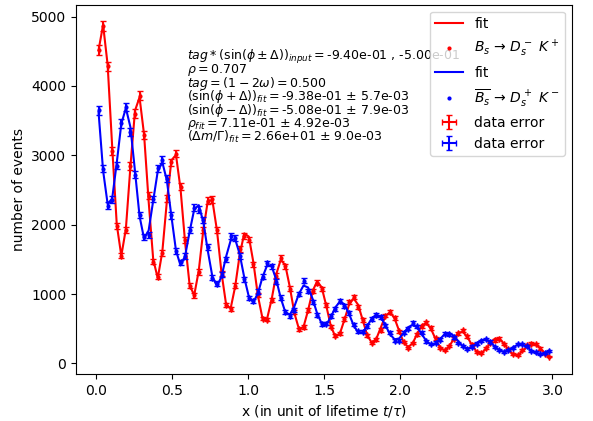} \qquad\\
\includegraphics[scale=0.55]{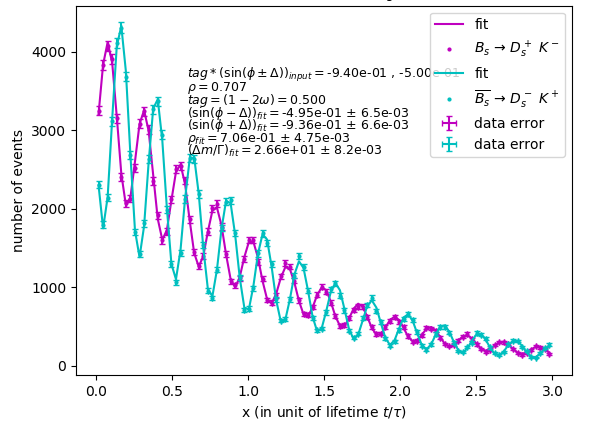} \qquad\\
\end{center}

Fig. 11: $\overline{B_s}(B_s) \to D_s^\pm K^\mp$ time-dependent distributions (upper and lower plots) with the statistics of 150 $ab^{-1}$ expected at FCCee.

\vskip 10pt

Note that this resolution could be further improved by using several other $B_s$ final states involving the same CKM phase, such as $D_s^{*\pm} K^\mp$ or $D_s^{\pm} K^{*\mp}$. In addition one can use other $D_s$ final states, such as $D_s^\pm \to \overline{K^{*0}}(K^{*0})K^\pm \to K^+K^-\pi^\pm$ and $D_s^\pm \to \phi \rho^\pm \to K^+K^-\pi^\pm\pi^0$. However, since several of these modes includes neutral particles, excellent electromagnetic calorimeter resolutions are crucial to reject the backgrounds.

\vskip 10pt

\begin{center}

\includegraphics[scale=0.5]{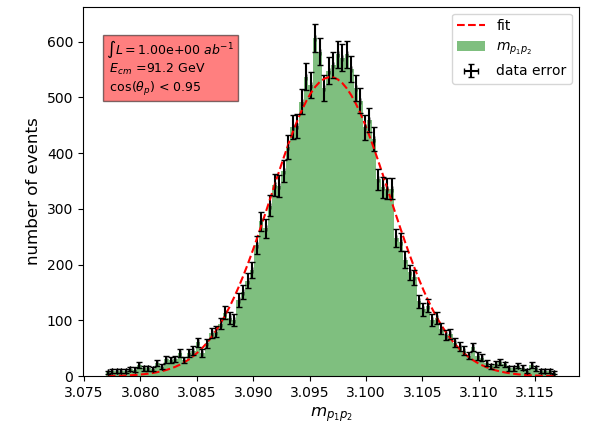} \qquad\\
\includegraphics[scale=0.5]{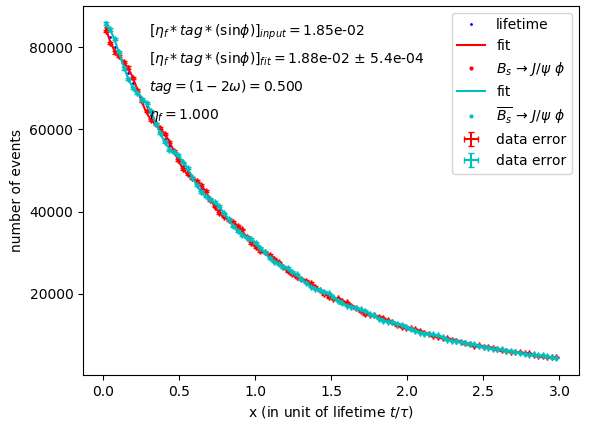} \qquad\\

\end{center}

Fig. 12: $\overline{B_s}(B_s) \to J/\psi \phi$ decay in $e^+ e^- \to Z \to \overline{B_s} (B_s) \to J/\psi \phi \to \mu^+ \mu^- K^+ K^-$. The upper plot shows the reconstructed $J/\psi$ mass. A resolution of $\sim 5.2$ MeV is observed. The two time-dependent distributions (equation (15)) are shown in the lower plot with the statistics of 150 $ab^{-1}$ expected at FCCee.

\vskip 10pt

We now focus on $\overline{B_s}(B_s )\to J/\psi \phi $ decaying to the 4-body final state $\mu^+\mu^- K^+K^-$. The geometric acceptance is very similar to  $D_s^\pm\pi^\mp \to K^+K^-\pi^+\pi^-$. The $B_s$ mass resolution is also very similar to $D_s^\pm\pi^\mp$. We show in Fig. 12 the $J/\psi$ mass resolution.

\noindent From the time-dependent fit (Fig. 12), $\sin\phi= \sin 2 \beta_s$ is measured. Thanks to the very hight statistics this lead to the resolution $\sigma(\beta_s) = 0.035^\circ $, while this angle is expected to be $\sim 1^\circ$. This sensitivity assumes that a full angular analysis is carried out since this 2-Vector final state includes $CP=+1$ and  $CP=-1$ contributions. If one does a simple time-dependent analysis as in Fig. 12, the amplitude of the interference is reduced by factor $1 -2f_\perp \simeq 0.5$, as already mentioned. This leads to $\sigma(\beta_s) = 0.07^\circ $. Interestingly, the impact of the present error on $f_\perp$ is negligible. Part of this reduced sensitivity can be mitigated by using additional $J/\psi$ final states such as $e^+e^-(\gamma)$ or other modes, such as $J/\psi \eta$, which is a pure $CP=+1$ eigenstate and does not necessitate an angular analysis. In this latter mode, excellent electromagnetic calorimeter is required to reject the backgrounds. In summary, the resolution of $\sigma(\beta_s) = 0.035^\circ$ mentioned above, is reacheable.

\section{ Conclusions}
In summary, we have shown that excellent resolutions on the angles $\alpha_s$ and $\beta_s$ are expected at FCCee  with $\sigma( \alpha_s) \simeq 0.4^\circ$ and  $\sigma( \beta_s) \simeq 0.035^\circ$. These angles are two angles of the unitarity triangle $UT_{sb}$ defined in equations~(\ref{eq:UT}) and Fig. 1. The angle directly measured by the mode $\overline{B_s}(B_s) \to D_s^\pm K^\mp$ is given by (\ref{eq:CKM_Bs_2bis}) $\phi_{D_sK} = \pi-(\alpha_s -\beta_s)$ and the mode $\overline{B_s}(B_s) \to J/\psi \phi$ gives a direct measurement of the angle (\ref{eq:phi_1}) $\phi_{J/\psi\phi} = \pi + 2\beta_s$. In a forthcoming paper to appear \cite{Aleksan-Oliver} it is shown that the modes $B^\pm \to \overline{D}^0(D^0)K^\pm \to K^+K^-K^\pm$ and $B^\pm \to \overline{D}^0(D^0)K^\pm \to K_s \pi^0 K^\pm$ enable one to measure directly $\phi_{D^0 K} = \pi + \gamma_s$. Interestingly, should unitarity hold, one has a simple relation between these observable phases, $-\phi_{D_sK} + \phi_{J/\psi\phi} + \phi_{D^0 K} = 0\ ({\rm mod}\ 2 \pi)$. All the angles of the ``flat'' unitarity triangle $UT_{sb}$ can thus be determined directly with high accuracy.

\vskip 15truemm

$$ \begin{tabular}{cccc}
\hline
&  & $\displaystyle {\mathrm {E_{cm} = 91.2\ GeV \ and\  \int L = 150 ab^{-1}}}$   &  \\

$\displaystyle {\mathrm {\sigma (e^+e^- \to Z )}} $ &
$\displaystyle {\mathrm {number}} $ &
$\displaystyle {\mathrm {f(Z\to \overline{B_s})}} $ &
$\displaystyle {\mathrm {Number \ of}} $ \\

$\displaystyle {\mathrm {nb}} $ & 
$\displaystyle {\mathrm {of \ Z} } $ &
$\displaystyle {\mathrm{}} $ &
$\displaystyle {\mathrm{produced \ \overline{B_s} }} $\\ 
\hline \hline 

$\displaystyle \sim 42.9$ &
$\displaystyle {\mathrm {\sim 6.4\ 10^{12}}}$ &
$\displaystyle {\mathrm {0.0159}} $ &
$\displaystyle \sim 1\ 10^{11} $\\ 
\hline
& & & \\

$\displaystyle {\mathrm {\overline{B_s}\ decay}} $ &
$\displaystyle {\mathrm {Decay}} $ & 
$\displaystyle {\mathrm {Final}} $ &
$\displaystyle {\mathrm {Number \ of}} $ \\

$\displaystyle {\mathrm {Mode}}  $ &
$\displaystyle {\mathrm {Mode} } $ &
$\displaystyle {\mathrm{State}} $ &
$\displaystyle {\mathrm{\overline{B_s} \ decays}} $ \\ 
\hline \hline
&  & $\displaystyle {\mathrm {nonCP \ eigenstates}}$   &  \\

$\displaystyle D_s^+\pi^-$ &
$\displaystyle {\mathrm {D_s^+\to \phi\pi}}$ &
$\displaystyle {\mathrm {K^+K^-\pi^+\pi^-}} $ &
$\displaystyle \sim 6.9\ 10^6 $\\ 
$\displaystyle D_s^+\pi^-$ &
$\displaystyle {\mathrm {D_s^+\to \phi\rho}}$ &
$\displaystyle {\mathrm {K^+K^-\pi^+\pi^-\pi^0}} $ &
$\displaystyle \sim 12.9\ 10^6 $\\ 
$\displaystyle D_s^+K^-$ &
$\displaystyle {\mathrm {D_s^+\to \phi\pi}}$ &
$\displaystyle {\mathrm {K^+K^-\pi^+K^-}}$ &
$\displaystyle \sim 5.2 \ 10^5 $\\ 
$\displaystyle D_s^+K^-$ &
$\displaystyle {\mathrm {D_s^+\to \phi\rho}}$ &
$\displaystyle {\mathrm {K^+K^-\pi^+K^-\pi^0}}$ &
$\displaystyle \sim 9.8 \ 10^5 $\\ 

$\displaystyle D^0\phi$ &
$\displaystyle {\mathrm {D^0\to K\pi}}$ &
$\displaystyle {\mathrm {K^-\pi^+K^+K^-}}$ &
$\displaystyle \sim 6.1 \ 10^4 $\\ 
$\displaystyle D^0\phi$ &
$\displaystyle {\mathrm {D^0\to K\rho}}$ &
$\displaystyle {\mathrm {K^-\pi^+K^+K^-\pi^0}}$ &
$\displaystyle \sim 1.7 \ 10^5 $\\ 
\hline
&  & $\displaystyle {\mathrm {CP \ eigenstates}}$   &  \\
$\displaystyle J/\psi\phi$ &
$\displaystyle {\mathrm {J/\psi\to \mu^+\mu^-}}$ &
$\displaystyle {\mathrm { \mu^+\mu^- K^+K^-}}$ &
$\displaystyle \sim 3.2 \ 10^6 $\\ 
$\displaystyle J/\psi\eta$ &
$\displaystyle {\mathrm {J/\psi\to \mu^+\mu^-}}$ &
$\displaystyle {\mathrm {\mu^+\mu^- \gamma\gamma}}$ &
$\displaystyle \sim9.6 \ 10^5 $\\ 

\hline

\end{tabular}   $$

Table 3: The expected number of produced $\overline{B_s}$ decays to specific decay modes at FCC-ee at a center of mass energy of 91 GeV over 5 years with 2 detectors. These numbers have to be multiplied by 2 when including $B_s$ decays. The branching fractions of the PDG~\cite{PDG 2020} have been used.

\newpage

\noindent {\Large \bf Appendix. Resolutions in the reconstructed flight\par \qquad \qquad distances}

\vskip3truemm

The resolutions in the reconstructed $B_s$ flight distances have been determined using Monte-Carlo events that were passed through a fast simulation of the tracking system of the experiment based on DELPHES \cite{de_Favereau_2014}. The PYTHIA8 Monte-Carlo generator was used to simulate the production of $b\overline{b}$ pairs and the decays of the $B_s$ mesons were performed with the EVTGEN program. The resulting charged particles were turned into simulated tracks using the "TrackCovariance" fast tracking software implemented in DELPHES. It relies on a description of
the tracker geometry. The vertex detector and the drift chamber of the IDEA detector \cite{Abada2019}, which provide resolutions similar to the ones given in Section 3.1, are used here. In this geometry, the central beampipe has an inner radius of $1.5$~cm and the innermost layer of the vertex detector is at $1.7$~cm of the beampipe center. The tracking software accounts for the finite detector resolution and for the multiple scattering in each
tracker layer and determines the (non diagonal) covariance matrix of the helix parameters that describe the trajectory of each charged particle. This matrix is then used to produce a smeared 5-parameters track, for each charged particle emitted within the angular acceptance of the tracker.\par

\vskip3truemm

A simple standalone code is used to fit a given set of tracks to a common vertex \cite{BedeschiCode}, under the assumption that the trajectories be perfect helices. When reconstructing the vertex corresponding to a given decay, a perfect "seeding" is assumed. For example, for the decay $B_s \to J/\psi \phi \to \mu \mu K K$, the vertex is reconstructed using the simulated tracks created from the muons and kaons that correspond to this decay. Once the $B_s$ decay vertex is reconstructed, the flight distance of the $B_s$ is taken to be the distance of this vertex to the
nominal interaction point \footnote{No attempt is made here to reconstruct the primary vertex of the event. However, it was checked that with a beam-spot constraint, the primary vertex can be reconstructed with a resolution of $4 - 5\ \mu m$ in $b\overline{b}$ events, leading to a negligible contribution to the resolution in the $B_s$ flight distance.}. The distribution of the difference between the reconstructed flight distance and the flight distance given by the Monte-Carlo information is approximately Gaussian, and its RMS can be used to define the resolution in the reconstructed flight distance. This resolution is shown in Fig. 13 as a function of the polar angle of the $B_s$, for the $B_s \to J/\psi \phi \to \mu \mu K K$ decay. As expected, better resolutions are obtained in the central region, since the tracks suffer less multiple scattering than in the forward region. The figure also shows the distribution of the normalised $\chi^2$ of the vertex fit, whose average is at one, as expected.\par

\vskip3truemm

\begin{center}

\includegraphics[scale=0.6]{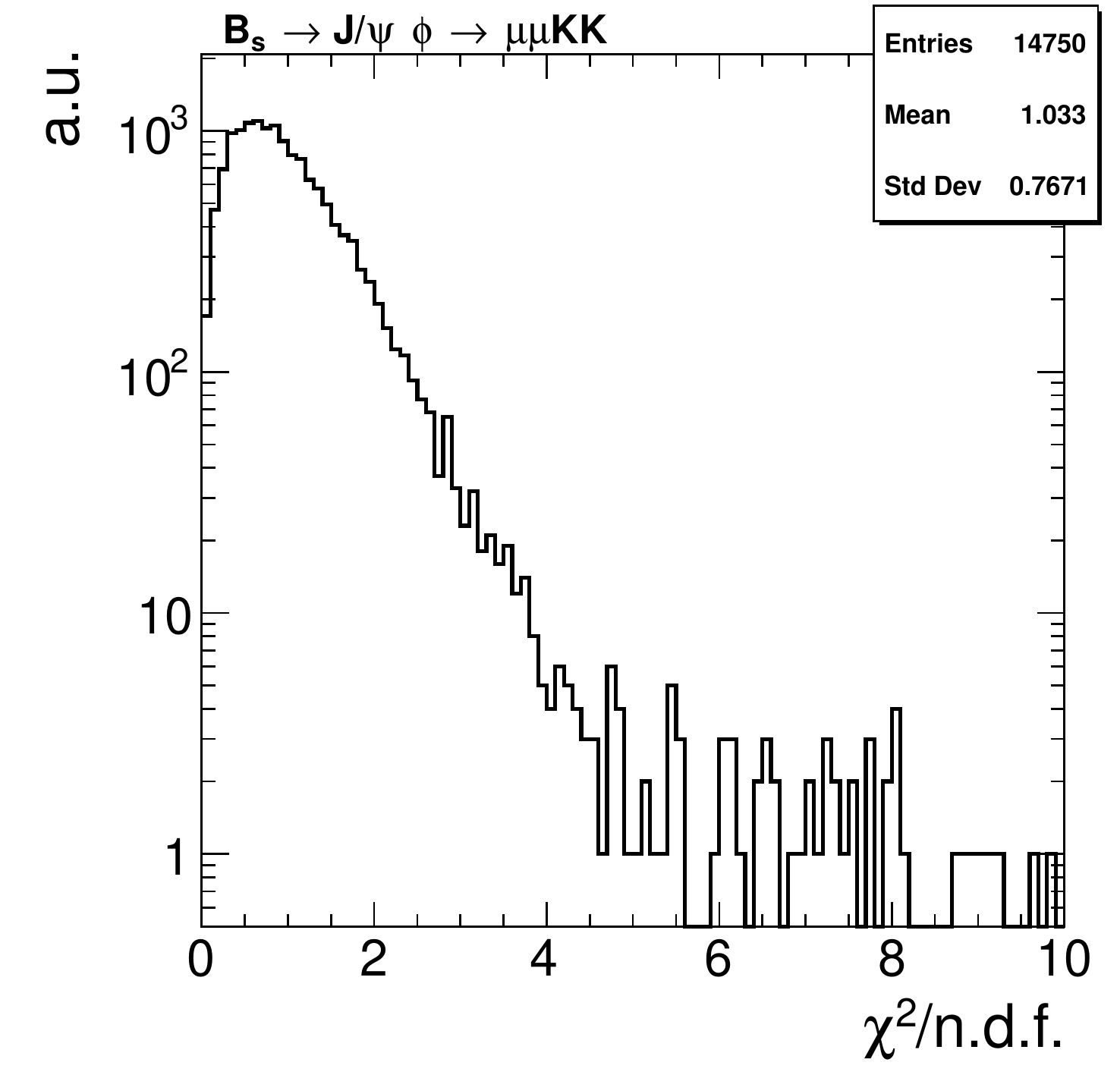} \includegraphics[scale=0.6]{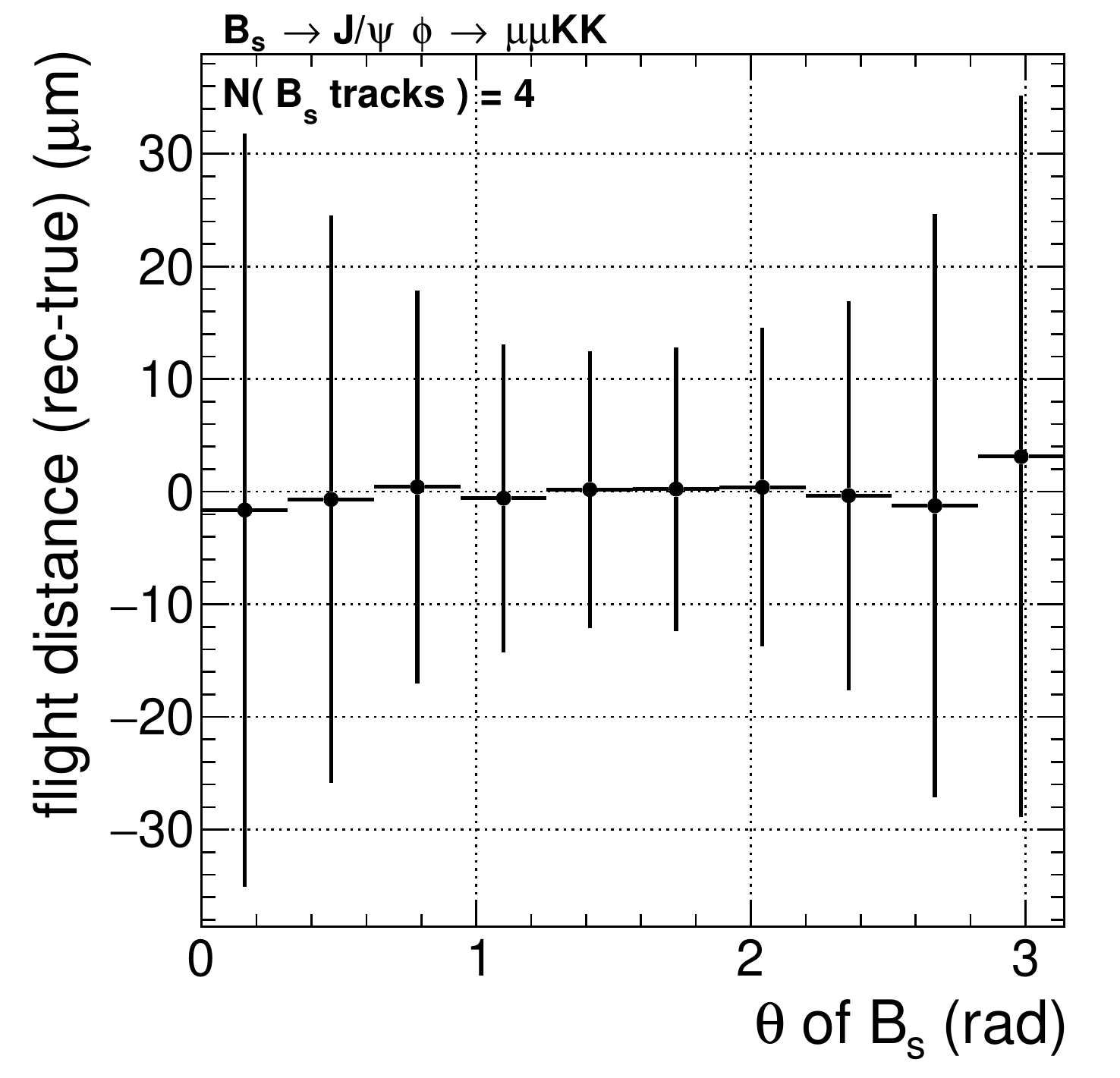} \qquad\\

\end{center}

Fig. 13. Upper plot: Distribution of the normalised $\chi^2$ of the fit of the $B_s \rightarrow J/\psi \phi \rightarrow \mu \mu KK$ vertex. Lower plot: Profile histogram of the difference between the reconstructed and the generated $B_s$ flight distance, as a function of the $B_s$ polar angle, for the $B_s \rightarrow J/\psi \phi \rightarrow \mu \mu KK$ decay. The error bars show the RMS of the distribution in each bin, hence the resolution in the flight distance.

\begin{center}

\includegraphics[scale=0.6]{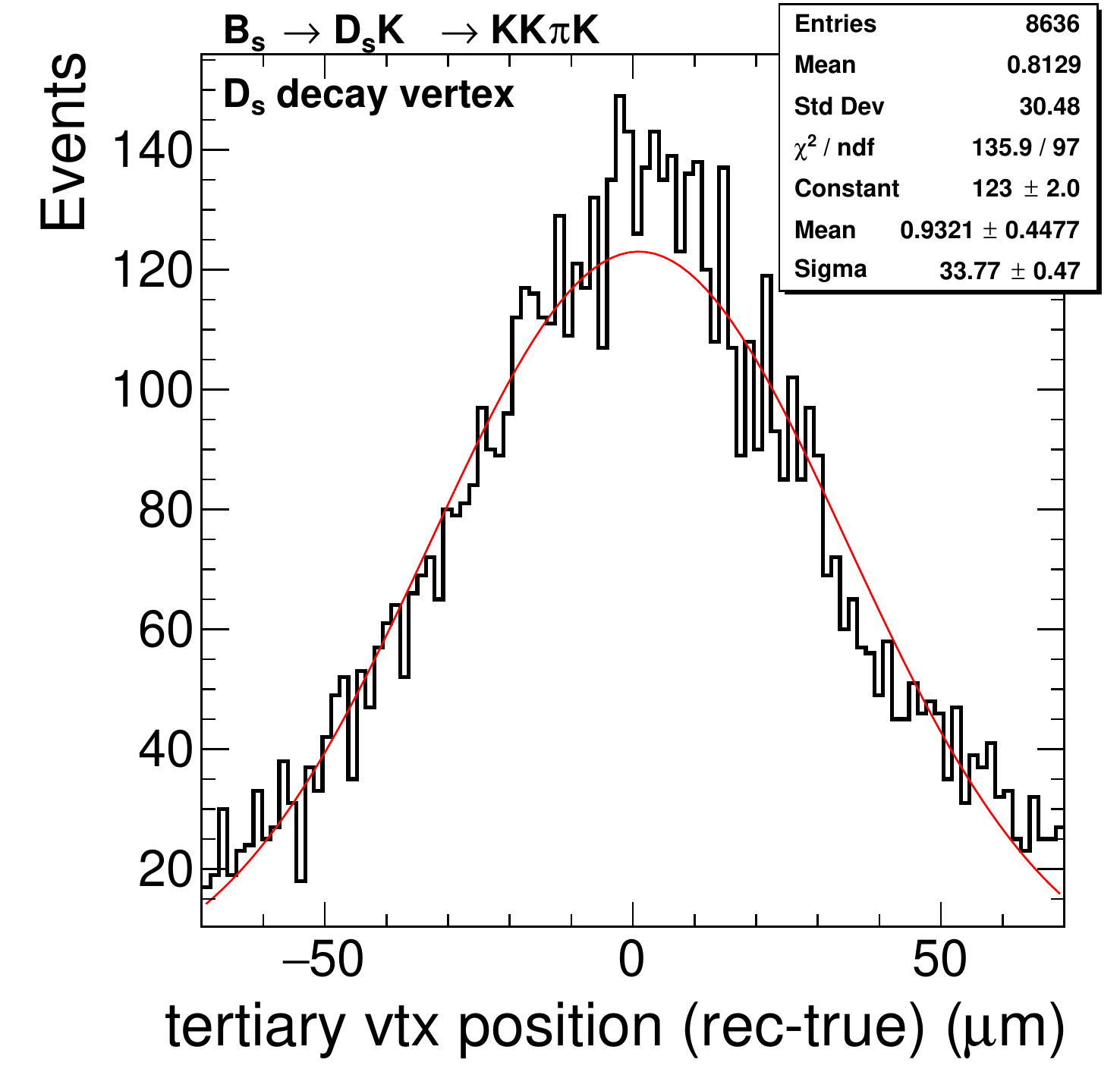} \includegraphics[scale=0.6]{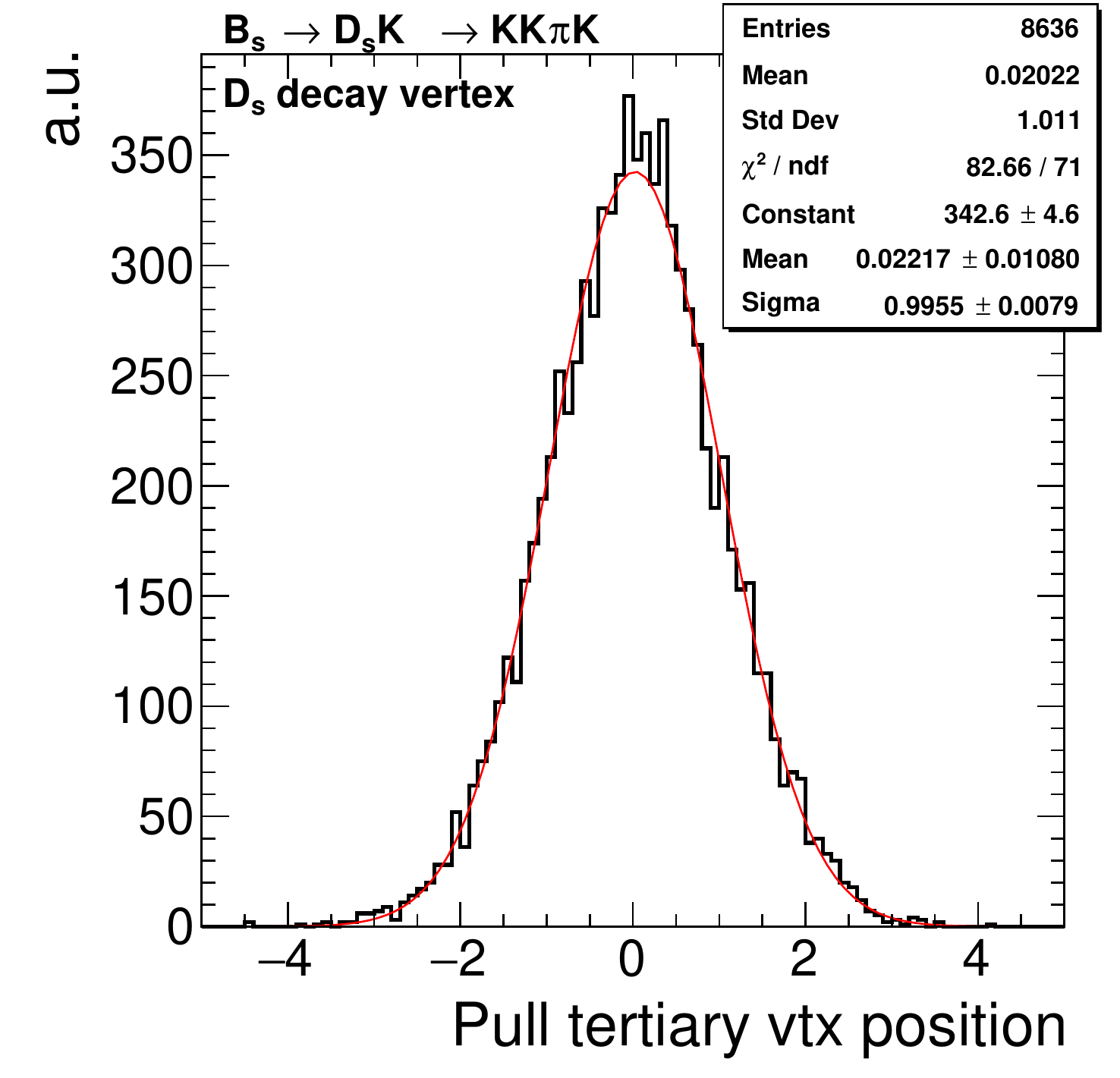} \qquad\\

\end{center}

Fig. 14. Upper plot: Resolution in the reconstructed position of the $D_s$ decay vertex in the decay $B_s \rightarrow D_s K \rightarrow K K \pi K$. Lower plot: pull of this reconstructed position. The red curves show the results of a Gaussian fit to the histograms. 

\vskip3truemm

The reconstruction of the $B_s$ decay vertex in $B_s \to D_s K$ is more complicated since $D_s$ does not decay promptly but leads to a tertiary vertex. In a first step, the $D_s \to KK\pi$ decay vertex is reconstructed, using the tracks corresponding to the three daughter particles of the $D_s$. When three tracks are reconstructed (about 95 \% of the events), the position of the $D_s$ vertex (the $3D$ distance between the vertex and the origin) is reconstructed with a resolution of about 33 $\mu$m, as shown in the upper panel of Fig. 14. The lower panel in the same figure illustrates the good quality of the fit: from the covariance matrix of the coordinates of the $D_s$ vertex, an uncertainty on the $D_s$ vertex position is computed, and is used to determine the corresponding pull, i.e. the difference between the reconstructed and the generated $D_s$ vertex positions divided by this uncertainty. As expected, the distribution of the pull is gaussian with a variance of unity. The momenta of the tracks corresponding to the $D_s$ daughters are then propagated to the $D_s$ decay vertex, such that their sum gives the three-momentum of the $D_s$ when it decays. From the $D_s$ momentum at its decay point and the coordinates of this decay vertex, a $D_s$ "pseudo-track" can be reconstructed, i.e. a set of 5-parameters that describe the $D_s$ helicoidal trajectory. In order to use this pseudo-track in the vertex fitter together with the track associated with the "bachelor" kaon (the $K$ from the $B_s \to D_s K$ decay) to reconstruct the $B_s$ decay vertex, the covariance matrix of the $D_s$ pseudo-track must be determined. This is done using a bootstrap method:\par

\vskip3truemm

$\bullet$ the parameters of the $D_s$ daughters' tracks are smeared according to their covariance matrices;\par

\vskip1truemm

$\bullet$ a $D_s$ vertex is re-fit using these smeared tracks, which are propagated to this vertex to define a $D_s$ momentum, and the 5-parameters of a $D_s$ pseudo-track;\par

\vskip1truemm

$\bullet$  the covariance matrix of the $D_s$ track parameters is obtained by statistical inference, over a large sample of $D_s$ pseudo-tracks obtained from smearing the $D_s$ daughter tracks many times.\par

\vskip3truemm

Figure 15 shows that, from the $D_s$ pseudo-track and the track of the bachelor $K$, the vertex fitter reconstructs the $B_s$ decay vertex with a resolution of about $18$~$\mu$m, and that the pull of this $B_s$ flight distance is distributed as expected.\par

\begin{center}

\includegraphics[scale=0.6]{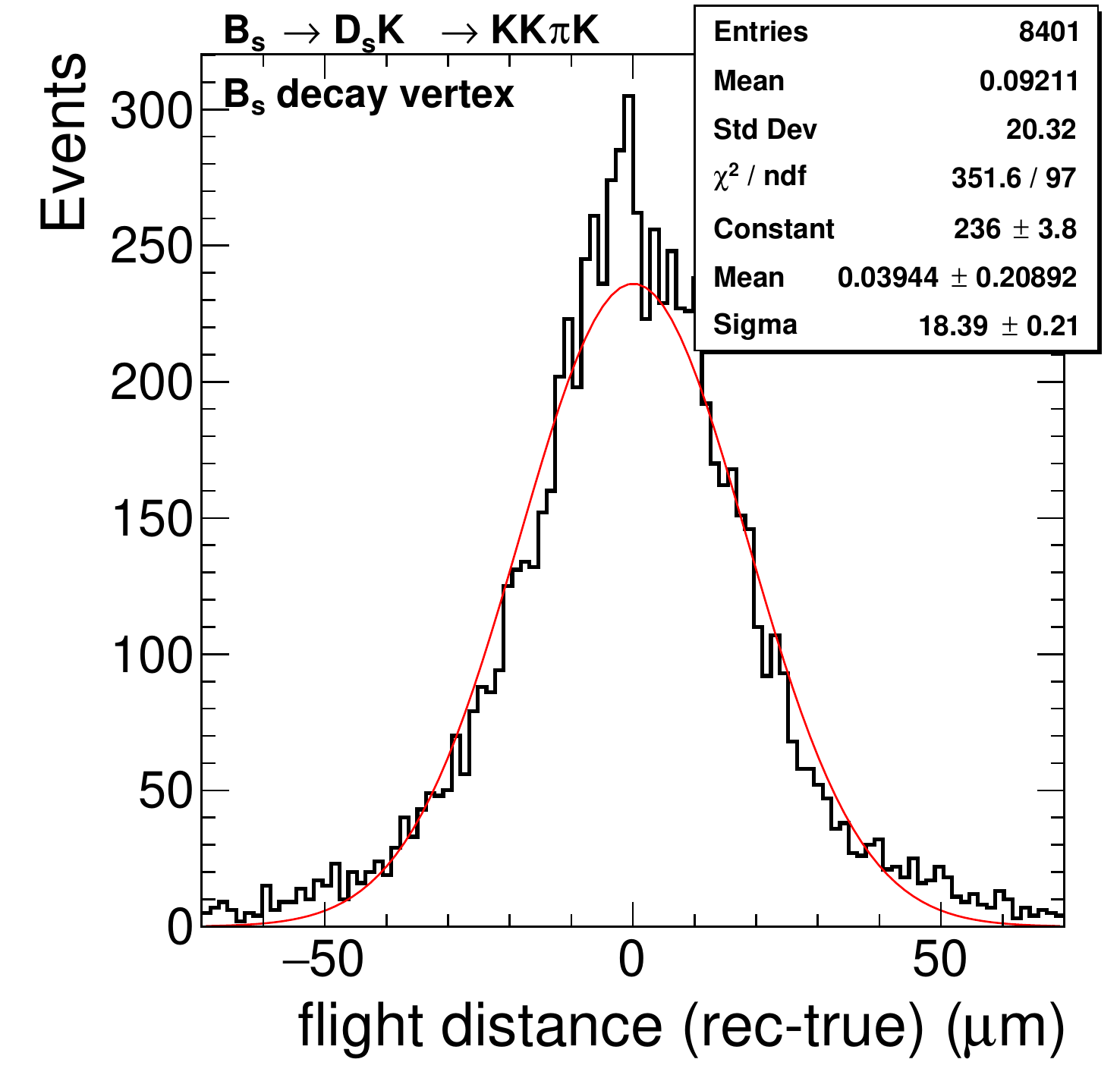} \includegraphics[scale=0.6]{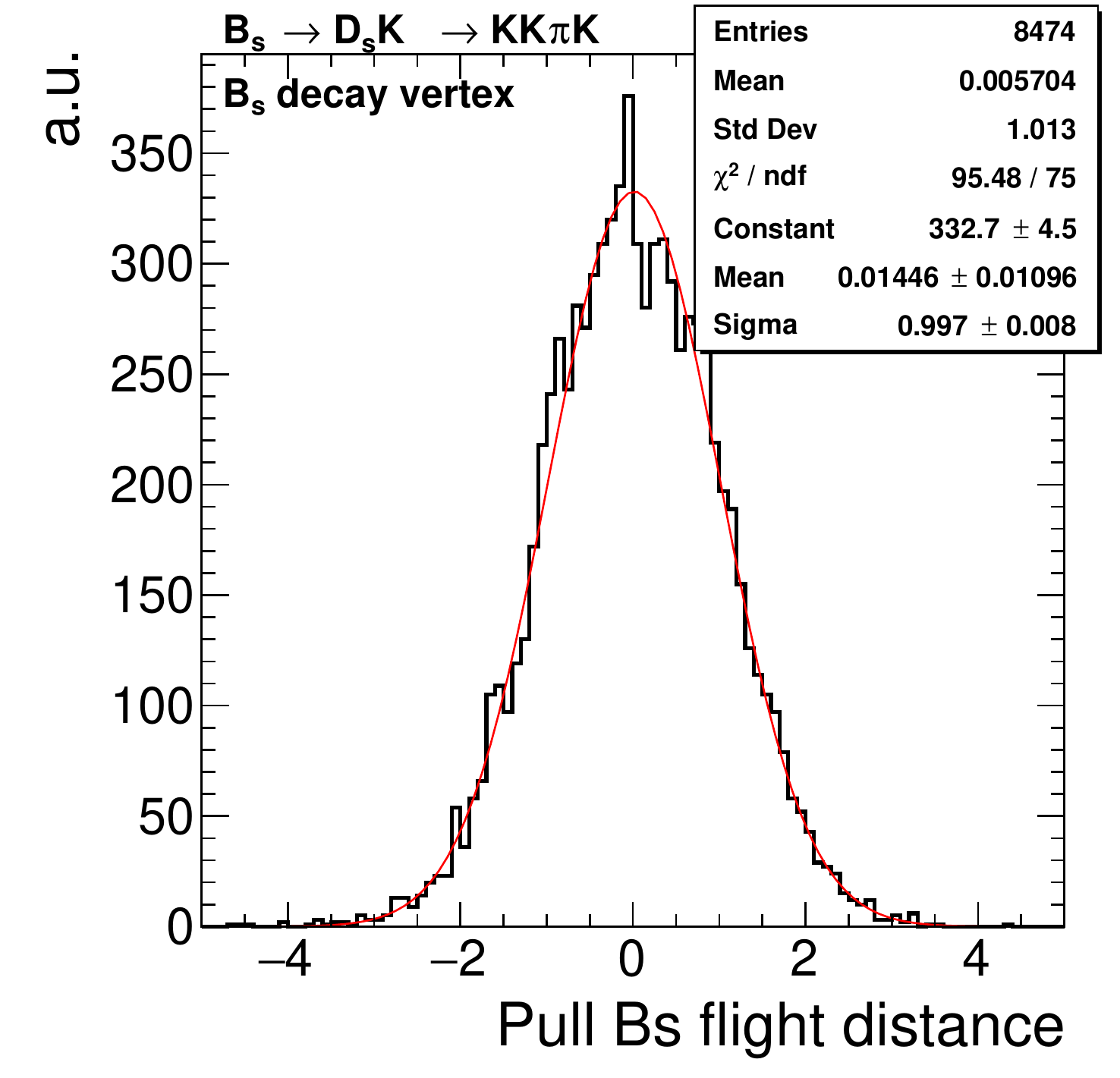} \qquad\\

\end{center}

Fig. 15. Upper plot: Resolution in the reconstructed flight distance of the $B_s$ in the decay $B_s \rightarrow D_s K \rightarrow K K \pi K$. Lower plot: pull of this flight distance. The red curves show the results of a Gaussian fit to the histograms.

\par

\end{document}

%% file: note_macro.tex
\def\Journal#1&#2&#3(#4){\unskip, #1~{\bf #2} (#4) #3}

\def\PTP{Prog.\ Theor.\ Phys.}

\def\ZPC{Z.\ Phys.\ C}

\def\U4S{\Upsilon (\mbox{4S})}

\def\b0b0{${\rm B^{o}\overline{B^o}}$}

\newcommand{\be}{\begin{equation}}
\newcommand{\ee}{\end{equation}}
\newcommand{\ba}{\begin{array}{c}}
\newcommand{\ea}{\end{array}}
\newcommand{\beqn}{\begin{eqnarray}}
\newcommand{\eeqn}{\end{eqnarray}}